\documentclass[aip,jmp,amsmath,amssymb,reprint]{revtex4-1}

\usepackage{hyperref}
\usepackage{epsfig}
\usepackage{graphicx}
\usepackage{subfigure}
\usepackage{latexsym}
\usepackage{color}
\usepackage{fullpage}
\usepackage{dcolumn}
\usepackage{bm}

\draft 

\begin{document}

\preprint{AIP/123-QED}


\newcommand {\JC}[1]{{\color{blue} #1}}

\title{Epitaxial graphene prepared by chemical vapor deposition on single crystal thin iridium films on sapphire}


\author{Chi Vo-Van}
 \email{Chi.Vo-Van@grenoble.cnrs.fr}
\affiliation{Institut N\'{e}el, CNRS \& Universite Joseph Fourier, BP166, 38042 Grenoble Cedex 9, France}

\author{Amina Kimouche}
\affiliation{Institut N\'{e}el, CNRS \& Universite Joseph Fourier, BP166, 38042 Grenoble Cedex 9, France}

\author{Antoine Reserbat-Plantey}

\affiliation{Institut N\'{e}el, CNRS \& Universite Joseph Fourier, BP166, 38042 Grenoble Cedex 9, France}

\author{Olivier Fruchart}

\affiliation{Institut N\'{e}el, CNRS \& Universite Joseph Fourier, BP166, 38042 Grenoble Cedex 9, France}

\author{Pascale Bayle-Guillemaud}

\affiliation{CEA-Grenoble, INAC/SP2M/LEMMA, 17 rue des Martyrs, 38054 Grenoble Cedex 9, France}

\author{Nedjma Bendiab}

\affiliation{Institut N\'{e}el, CNRS \& Universite Joseph Fourier, BP166, 38042 Grenoble Cedex 9, France}

\author{Johann Coraux}
 \email{Johann.Coraux@grenoble.cnrs.fr}
\affiliation{Institut N\'{e}el, CNRS \& Universite Joseph Fourier, BP166, 38042 Grenoble Cedex 9, France}


\date{\today}

\begin{abstract}
Uniform single layer graphene was grown on single-crystal Ir films a few nanometers thick which were prepared by pulsed laser deposition on sapphire wafers. These graphene layers have a single crystallographic orientation and a very low density of defects, as shown by diffraction, scanning tunnelling microscopy, and Raman spectroscopy. Their structural quality is as high as that of graphene produced on Ir bulk single crystals, \textit{i.e.} much higher than on metal thin films used so far.
\end{abstract}

\maketitle 

Since 2004, there has been an increasing effort in developing efficient methods for preparing graphene, mostly motivated by the prospect of applications. By chemical vapour deposition (CVD) on transition metal surfaces large-area few-layer of epitaxial graphene can be obtained and then transferred in principle to any support. CVD of graphene has been long performed on bulk transition metal single crystals under ultra-high vacuum (UHV).\cite{Wintterlin2009} In 2008, a more versatile method was introduced, using polycristalline Ni films on Si wafers as substrates for CVD, which was operated closer to atmospheric pressure.\cite{Reina2009,Kim2009} However, due to the relatively high solubility of carbon in Ni, the precise control of the graphene layer thickness down to single layer turned out tedious.\cite{Kim2009} This was partly circumvented by employing Cu films, for which C solubility is much lower;\cite{Li2009} yet, multilayer graphene could not be avoided at the location where Cu grain boundaries cross the surface. Intrinsic limitations to the quality of graphene are also imposed by  substrate roughening at the high temperature of CVD.\cite{Wofford2010} Moreover polycrystalline metal films impose twinned domains in graphene, with length scale in the range of the metal grain size (typically a few 10~$\mu$m). Since graphene's properties depend on the number of layers, and charge carrier scattering is believed to take place at grain boundaries or substrate-induced rippling –both altering charge carrier's mobility, so far limited to a few 10$^3$~cm$^2$V$^{-1}$s$^{-1}$ for CVD grown graphene–, higher and better defined performances are expected for higher quality graphene. 

Employing high quality thin metal films onto which a continuous graphene sheet with single crystallographic orientation could be achieved is thus a promising route, and was reported only very recently on Co,\cite{Ago2010} Ru,\cite{Sutter2010}, and Ni.\cite{Iwasaki2010} In this Letter we report the preparation of single crystalline Ir nanometer-thick films on sapphire, and of large-area, high-quality single layer graphene on top. Iridium is one of the low C solubility metals onto which pure single layer graphene can be prepared, which is favorable for the achievement of high quality graphene; it is also known to weakly interact with graphene, which allows for the fine manipulation of graphene's electronic band structure, for instance the opening of a band gap at the Dirac point.\cite{Balog2010} Graphene on Ir thin films on sapphire is thus a model system for multi-technique investigations, including \textit{ex situ} (which is not so convenient with bulk single crystals), of the properties of graphene contacted to a metal.

\begin{figure}
\includegraphics[width=80mm]{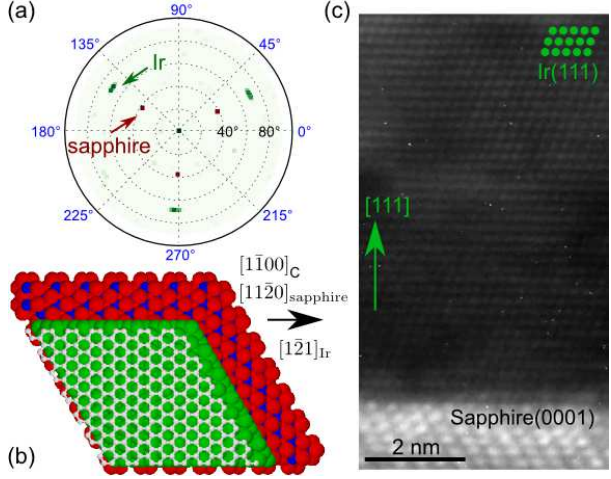}
\caption{\label{fig:1} (a) X-ray pole figures (logarithm of the scattered intensity) along a $[10\bar{1}4]$ direction of sapphire (red) and a $[111]$ direction of Ir (10~nm-thick, grown at 700~K and annealed at 1100~K, green) on the same spherical coordinate system (Azimuthal and scattering angles displayed in blue and black respectively). The Ir Bragg peaks are broader than the sapphire ones, indicative of some spread in the epitaxial relationship. (b) Ball-model of graphene/Ir/sapphire (white: graphene, green: Ir, red/blue: sapphire). (c) TEM cross section of a 9~nm Ir film on sapphire, along the $\left\lbrack1\bar{1}0\right\rbrack$ azimuth. Green dots highlight the crystallographic structure of Ir.}
\end{figure}

Sapphire [$\alpha$-Al$_2$O$_3$(0001)] wafers (Roditi Ltd.), cut in 6.5$\times$8.5~mm$^2$ pieces, were used as substrates. Iridium was grown by pulsed laser deposition (0.1-1~J$\times$cm$^{-2}$, 10~ns, and 10~Hz pulse fluence, duration, and repeat frequency) at a rate of 0.1~nm$\times$min$^{-1}$ in a UHV chamber (base pressure 5$\times$10$^{-11}$~mbar). As discussed later, we find that growth at 700~K and 30~min annealing to 1100~K yield the best quality. CVD was performed in a second UHV chamber connected to the first one (base pressure 10$^{-10}$~mbar), with ethylene as a carbon precursor, which was brought in the vicinity of the sample via a dosing tube ensuring a local partial pressure higher than that in the chamber (typically 10$^{-8}$~mbar for the latter). CVD during 10~min above 950~K ensured a graphene coverage in excess of 95~\%. Scanning tunnelling microscopy (STM) and reflection high-energy electron diffraction (RHEED) were performed \textit{in situ}, temperatures were measured with a pyrometer. X-ray diffraction, transmission electron microscopy (TEM, 400~kV), and Raman spectroscopy (WITec alpha500, 532~nm) were conducted \textit{ex situ}.

We first address the crystalline structure of the Ir thin films. Figure~\ref{fig:1}a shows X-ray pole figures of sapphire and of the Ir thin film (10~nm). The three-fold symmetry and Bragg diffraction angles prove that the Ir is (111)-textured and has a well-defined in-plane epitaxial relationship with sapphire, $\left\langle11\bar{2}0\right\rangle_\mathrm{sapphire} \parallel \left\langle1\bar{2}1\right\rangle_\mathrm{Ir}$ (Fig.~\ref{fig:1}b), with a full-width at half-maximum (FWHM) spread of 1$^\circ$ as derived from azimuthal angle scans. Before the 1100~K annealing step, X-ray pole figures and STM (see later) evidence 180$^\circ$ twins along [111]. Consistently, TEM cross-sections (such as in Fig.~\ref{fig:1}c) confirm that a single orientation is obtained following the annealing step, and that the Ir(111) films are single-crystalline.

\begin{figure}
\includegraphics[width=80mm]{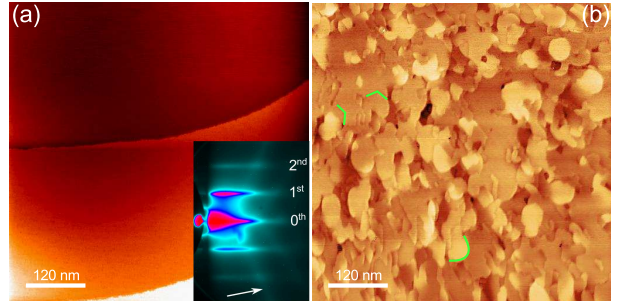}
\caption{\label{fig:2} STM topographs of 10~nm-thick Ir(111) films on sapphire, after (a) a 1100~K and (b) a 900~K annealing. Green curves (b) highlight two kinds of surface features (see text). The inset in (a) is a RHEED pattern (20~kV, $[1\bar{1}0]$ azimuth) showing the 0$^\mathrm{th}$, 1$^\mathrm{st}$ and 2$^\mathrm{nd}$ order streaks of Ir(111) and Kikuchi lines (white arrow).}
\end{figure}

We then characterize the surface quality of Ir(111). Streaky RHEED patterns (Fig.~\ref{fig:2}a, inset) suggest that the Ir(111) surface is atomically smooth at the scale of a few 10~nm. Kikuchi lines are further evidences for the quality of the surface. The patterns correspond to a single crystalline surface, consistent with the volume characterization (X-ray diffraction and TEM). STM after annealing reveals atomically smooth terraces, separated by atomic step edges, and $\simeq$300~nm-wide (Fig.~\ref{fig:2}a), a width directly related to the miscut of the wafer. Comparably smooth and wide terraces are obtained on bulk single crystals only after ion bombardment and high temperature flash ($\simeq$1600~K for Ir) which promotes the formation of metal step bunches, here absent. The very thin metal films do not dewet upon annealing, probably thanks to the small lattice mismatch between Ir(111) and sapphire(0001) and to the high melting temperature of Ir. The influence of the annealing step is obvious when comparing Figs.~\ref{fig:2}a and b. For too mild annealing (900~K), the film surface exhibits curved features, which are atomic step edges, as well as lines which are 120$^\circ$ rotated one with respect to the other. We speculate that these are the surface traces of grain boundaries between 180$^\circ$ in-plane twins which are detected on such films with X-ray pole figures (not shown). Iridium is fully relaxed to its bulk lattice parameter as evidenced by the formation of a network of misfit dislocations at the Ir/sapphire interface (STM data not shown).

\begin{figure}
\includegraphics[width=75mm]{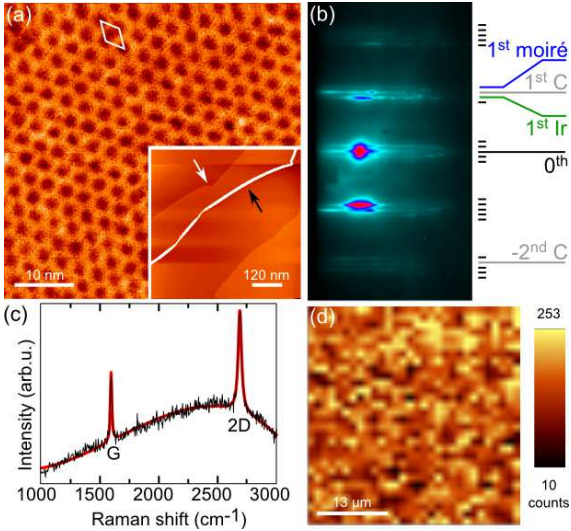}
\caption{\label{fig:3} Single graphene layer grown at 1300~K on the Ir(111) film: (a) STM topographs evidencing the moir\'{e} (white rhombus), and showing (inset) graphene wrinkles (black arrow) as well as reshaping of the underneath Ir step edges (white arrow). (b) RHEED pattern of graphene/Ir(111) (10~kV, $[1\bar{1}0]$ azimuth). First order Ir, graphene, and moir\'{e} streaks are highlighted as well as the zeroth order streak and one second order moir\'{e} streak. (c) Background corrected Raman spectrum showing G and 2D speaks and (d) 30$\times$30~$\mu$m$^2$ map of the G band intensity (laser wavelength: 532~nm, power: 1.7~mW$\mu$m$^{-2}$).}
\end{figure}

Graphene growth on the Ir(111) thin films proceeds like on bulk single crystals.\cite{Coraux2009} Small-angle twinned domains, having an extension between a few 10 to a few 100~nm, are found for growth performed at 1100~K, while a single crystallographic orientation, corresponding to the typical graphene/Ir(111) moir\'{e} (Fig.~\ref{fig:3}a), is found at 1400~K. This orientation is $\left\langle1\bar{1}00\right\rangle_\mathrm{C} \parallel \left\langle1\bar{2}1\right\rangle_\mathrm{Ir}$, as shown by RHEED. Graphene rods are only found along the $[1\bar{1}0]$ azimuth of Ir(111). In this orientation, also superstructure rods from the moir\'{e} are observed (Fig.~\ref{fig:3}b). The distances from the Ir, graphene, and moir\'{e} streaks to the central one give an estimate for the in-plane lattice parameters of graphene and the moir\'{e}, 0.246$\pm$0.001~nm and 2.634$\pm$0.014~nm respectively, which is similar to the situation on bulk single crystals.\cite{NDiaye2008} STM topographs display the typical wrinkles\cite{Kim2009} and Ir step edge reshaping\cite{Coraux2009} (black and white arrows in the inset of Fig.~\ref{fig:3}a). Neither STM nor RHEED detect multilayer graphene. The Raman spectra presented in Fig.~\ref{fig:3}c is averaged over 20$\times$20~$\mu$m$^2$. It shows the characteristic features for single-layer graphene, \textit{i.e.} the G and 2D peaks, whose frequency shifts are 1594~cm$^{-1}$ (FWMH: 7~cm$^{-1}$) and 2688~cm$^{-1}$ (FWMH: 17~cm$^{-1}$) respectively. As compared to graphene on other supports (epitaxially grown or exfoliated), G and 2D band positions are both shifted to higher frequency and their FWHM are extremely sharp, most prominently for the G band. These observations indicate a charge transfer between graphene and its Ir support,\cite{Ferrari2008} the sign of the shift suggesting p-doping of graphene, which agrees with photoemission spectroscopy.\cite{Pletikosic2009} The absence of a D band implies that graphene is quasi-defect-free, consistent with the STM data. The narrow width of the G band rules out the presence of amorphous carbon. The strong background observed for all spectra is due to Ir luminescence. Previous attempts in measuring Raman spectra for graphene on Ir bulk single-crystals remained unsuccessful, which leads us to the conclusion that the luminescence from bulk Ir, unlike in the case of Ir thin films, overwhelms the graphene signature. Figure~\ref{fig:3}d shows a large-area spatial map of the Raman G band intensity. This map is remarkably homogeneous with no graphene-free region and is representative of maps aquired at other locations on the sample. The observed fluctations are artifacts originating from low signal-to-noise ratio due to the strong Ir background.

In conclusion, we have shown that extremely high quality, large area, single layer graphene can be prepared by UHV CVD on few nanometer-thick single crystalline Ir(111) films grown on sapphire wafers. A further step in the production of graphene would consist in transferring the graphene to another support.

We thank Philippe David and Val\'{e}rie Guisset for technical support, Luc Ort\'{e}ga for X-ray pole figures measurement, Violaine Salvador for the TEM sample preparation and measurements, Zahid Ishaque for STM measurements. CVV and AK acknowledges financial support from the Nanosciences Fondation and EU contract No. NMP3-SL-2010-246073 (GRENADA).

\end{document}